# Electric-field-assisted phase switching for crystal phase quantum dot fabrication in GaAs nanowires

Qiang Yu,[1] Khakimjon Saidov,[2] Ivan Erofeev,[2] Khalil Hassebi,[1] Chen Wei,[1] Charles Renard,[1] Laetitia Vincent,[1] Frank Glas,[1] Utkur Mirsaidov,[2*] Federico Panciera[1*]

[1] Université Paris-Saclay, CNRS, Centre de Nanosciences et de Nanotechnologies, 91120, Palaiseau, France

[2] Centre for BioImaging Sciences, Department of Biological Sciences and Physics, National University of Singapore, Singapore 117557

**Abstract:** The occurrence of several crystal phases within nanostructures of a single material presents both challenges and opportunities. While unintended phase mixing can degrade optoelectronic performances, deliberate control of polytypism enables novel heterostructures with unique quantum properties, crystal phase quantum dots (CPQDs). However, since tailoring the formation of CPQDs is difficult, applications remain scarce. Here, we demonstrate electric-field-driven crystal phase switching in GaAs nanowires during vapor-liquid-solid growth, enabling precise control of crystal phases and creations of CPQDs with monolayer precision. Nanowires are epitaxially grown on custom-made silicon micro-substrates using chemical vapor deposition within an in-situ TEM. Real-time imaging reveals that the electric field switches instantaneously the crystal phase between zinc blende and wurtzite, creating atomically sharp interfaces. Numerical simulations are developed to investigate the impact of the electric field on the catalyst droplet geometry, which largely governs the crystal phase. This constitutes a key progress towards unlocking the potential of CPQDs.

The vapor-liquid-solid (VLS) method is widely used for the growth of semiconductor nanowires (NWs). While bulk semiconductors such as GaAs generally exist in only one crystal phase, at the nanoscale, surface energy contributions may stabilize metastable phases that are inaccessible in bulk materials. For instance, GaAs NWs grown via VLS may adopt the thermodynamically stable cubic zinc blende (ZB) structure or the metastable hexagonal wurtzite (WZ) structure, which often results in a mixed-phase structure.[1] Remarkably, the staggered type II alignment of the valence and conduction bands of the two phases allows small sections of one phase within the other to confine either electrons or holes.[2,3] Provided that their radial (NW diameter) and axial (heights of single phase segments) dimensions are small enough, such heterostructures truly constitute crystal phase quantum dots (CPQDs). They feature, for example, sharp and intense spectral lines and single photon emission[2,4–6]. More importantly, contrary to compositional heterojunctions, CPQDs may have atomically abrupt ZB-WZ interfaces and benefit from the very low mismatch between the two phases[7]. This makes them strong candidates for several potential applications in photonics[2,8] and quantum computing[9,10].

However, their technological application has been severely limited, mainly by the difficulty of controlling their formation[4,5,11]. Thus, the investigation of the optical properties of these structures had to rely on accidentally formed CPQDs[2,4,8]. A key challenge is to achieve precise control of phase switching while maintaining high crystal quality.

One of the central parameters controlling the crystal phase, namely the contact angle between the liquid droplet and the NW, was identified theoretically early on[12]. Only recently, thanks to *in situ* transmission electron microscope (TEM)[13,14], could researchers shed new light on the phase selection mechanism and demonstrate how to achieve structural control[15]. *In situ* TEM provides atomic resolution imaging and captures the growth dynamics and the effects of changing growth parameters in real time[16–19]. It was shown that changes in growth conditions could modify the catalyst droplet volume and contact angle and result in phase switches[13,14,20]. In GaAs, phase changes were found to occur at two critical contact angles of 100° and 125°[13], with the intermediate range corresponding to the WZ phase and contact angles larger than 125° or smaller than 100° to ZB. In this work, we focus on the critical angle of 125°. Since this transition does not involve NW tapering, the NW diameter may be kept uniform during growth.

Until now, the only way to change the contact angle on purpose was to modify the droplet volume by acting on the growth conditions, such as vapor fluxes or temperature[3,21]. With metalorganic chemical vapor deposition (MOCVD) or molecular beam epitaxy (MBE), changes in droplet volume typically occur on a time scale much longer than the formation of a single monolayer (ML) of the NW. This tends to produce a mixed-phase segment at the interface between WZ and ZB regions. This often leads to the formation of a mixed-phase segment at the interface between the WZ and ZB regions, which can only be avoided through precise and intricate manipulation of the growth parameters.[3]. A method enabling a rapid modulation of the contact angle and phases, on the timescale of the formation of a single ML, while preserving straight sidewall, is thus highly desirable. Such a method was explored for Au-catalyzed Si NWs[22]. By submitting the NW/droplet system to an electric field (E-field) in an *in situ* TEM setup, it was demonstrated that the contact angle can be modulated quasi-instantaneously by changing the E-field amplitude[22] without changing growth temperature and flux. In these experiments, the structure remains cubic diamond.

Here, we use an E-field to trigger the phase changes quasi-instantenously in GaAs NWs grown by Au-catalyzed MOCVD in an *in situ* TEM without changing any other growth parameters such as temperature or gas pressure, and, in doing so, create CPQDs with the ultimate ML precision. This method constitutes a breakthrough in the fabrication of CPQDs and, more generally, in the synthesis of epitaxial semiconducting quantum dots.

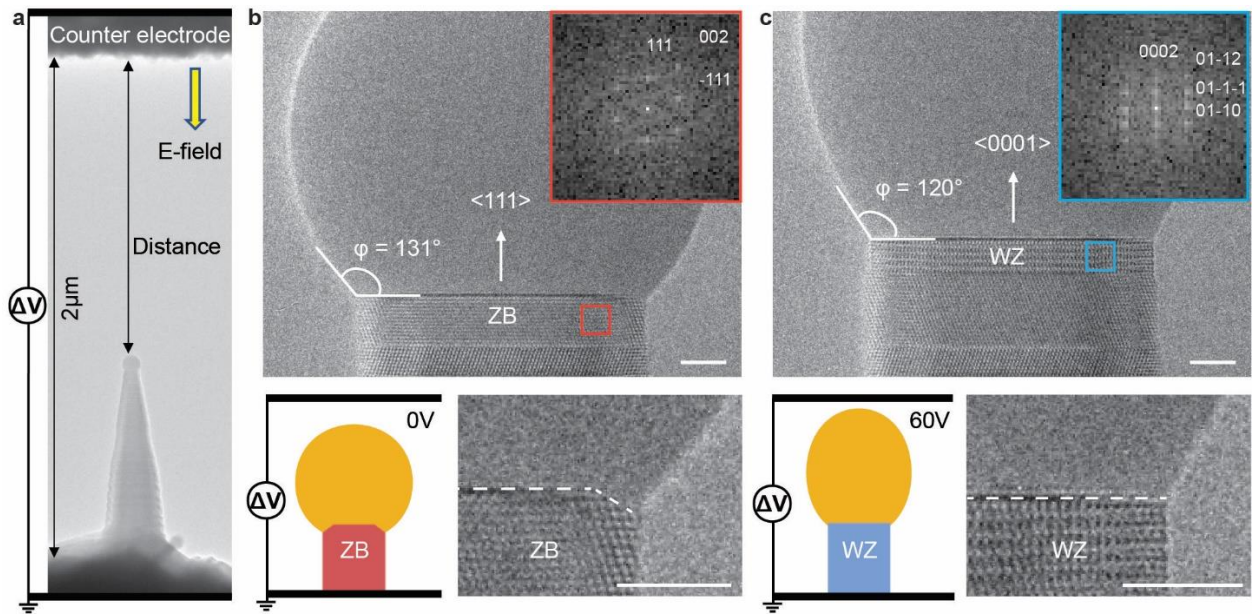

**Figure 1.** Crystal phase changes induced by an E-field. **a** Low magnification TEM image showing an isolated GaAs NW growing on a microfabricated cantilever. The bottom side is grounded, while the top part, the counter-electrode, can be set at different biases to deform the droplet. **b** Before the application of an E-field, the droplet on top of the NW (different from a) has a contact angle φ larger than 130°. The lower inset highlights the truncation at the TPL and shows that the crystal phase is ZB, as confirmed by the fast Fourier transform (FFT) of the area within blue square (upper inset). **c** Same NW under a 60 V bias on the counter-electrode. The contact angle is reduced to 120°, the truncation at the TPL disappears (inset) and the crystal phase is WZ. Scale bars: 5 nm.

To explore the feasibility of manipulating the crystal phase via an E-field, Au-catalyzed GaAs NWs are grown by MOCVD on microfabricated Si cantilevers[23] specially designed to fit into our *in situ* TEM (Fig. 1a; details of cantilever structure are given in Supporting information Fig. S1). The cantilevers provide a crystalline substrate for epitaxial growth in a <111> direction and a counter-electrode, which can be biased for applying the E-field along this direction. In Fig. 1b, recorded under zero bias, the NW grows under a spherical cap droplet exhibiting a contact angle of 131°. This results in the formation of a ZB segment with the characteristic oscillating truncation at the triple phases line (TPL)[13,24,25]. Upon applying a 60 V bias (Fig. 1c), the catalyst droplet elongates along the E-field direction without changing volume, and the contact angle is reduced to 120°. At the same time, the truncation at the TPL is suppressed, and the WZ phase appears. The coupling between interface morphology and crystal structure demonstrated here is similar to that observed in previous experiments where growth conditions (mainly fluxes) were manipulated to change the droplet volume[13,14,21]. These similarities suggest that the mechanism of phase selection is unaffected by the application of an E-field. As in previous experiments[22], we find that the deformation of the droplet is independent of the polarity of the E-field. This can be understood simply as resulting from a reversal of the distribution of the positive and negative surface charges upon bias reversal.

A clear advantage of using an E-field compared to altering the growth conditions lies in its ability to abruptly change the contact angle to values far from the critical angle. The contact angle is typically measured when the NW is observed along a ZB <110> zone axis, a convention widely adopted in literature and followed throughout this paper. However, a tomography study of Ge NWs[26] has shown that, for a given droplet volume, a range of contact angles—spanning up to 15 degrees—can coexist along TPL. This effect, though less pronounced, is also observed in GaAs

NWs. As a result, when the contact angle is close to the critical value, monolayer nucleation may occur at different points where the local contact angle is below or above the critical angle, which may lead to the formation of mixed phases. The abrupt change in contact angle induced by the E-field effectively prevents the nucleation at values close to the critical one and thus the formation of such mixed-phase segments and more generally growth in conditions that remain difficult to control and poorly understood[27].

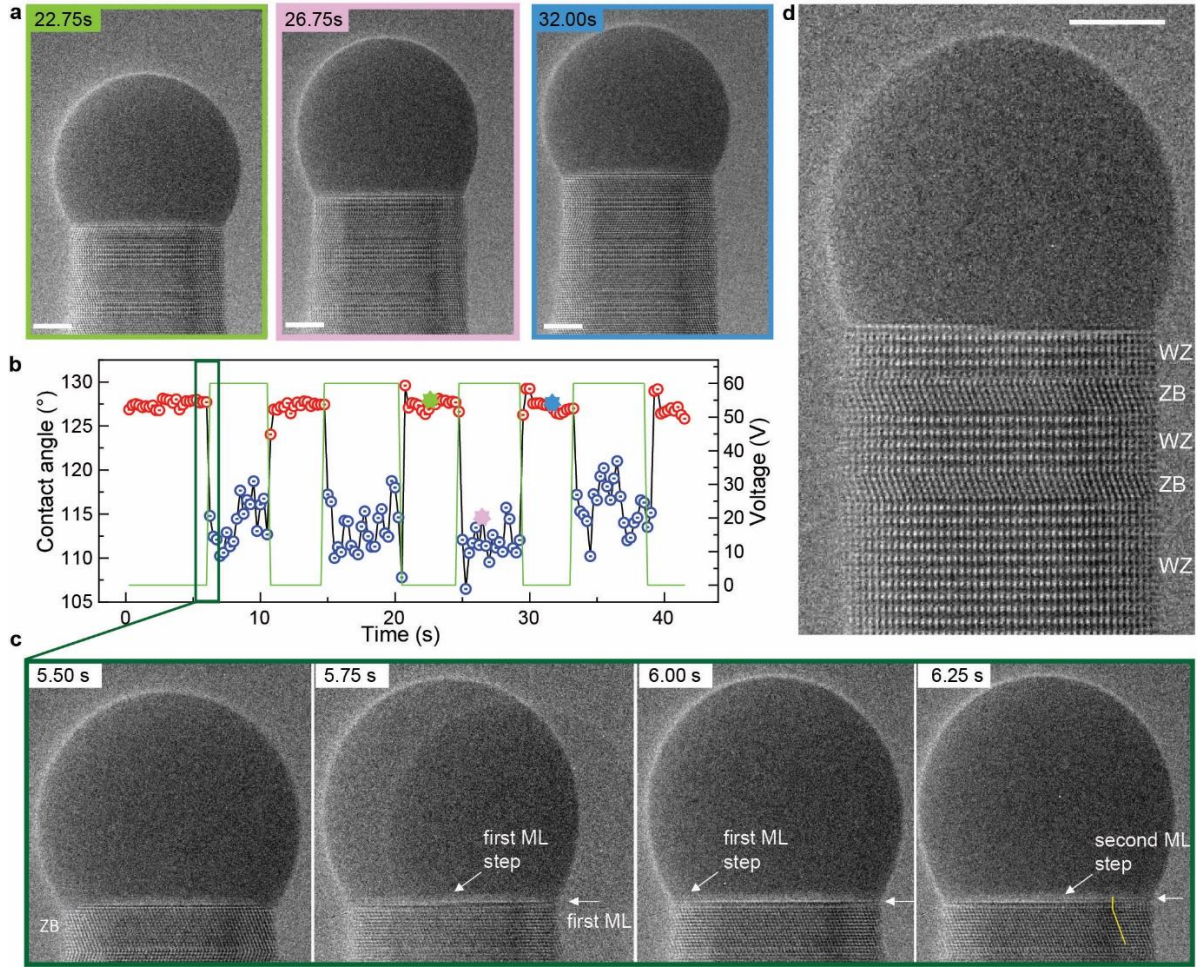

**Figure 2:** Fabrication of multiple CPQDs. **a** Sequence of TEM images extracted from Supporting Video SV1 recorded during the growth of a GaAs NW under E-field. The bias is modulated between 0 and 60 V, which results in alternating WZ and ZB segments. (since the contact angle is close to the critical angle, there are stacking faults) **b** Correlation between measured droplet contact angle (symbols) and applied voltage (green curve). Each data point corresponds to a video frame (collected at 4 frames per second) and is color-coded according to the crystal phase growing at that time (red for ZB, blue for WZ). The green, pink and blue stars correspond to the three frames in **a**. **c** Four successive frames of supporting video SV1. At 5.50 s, a ZB segment is growing. In the 5.75 s frame (which integrates over times before and after E-field application, hence the double image of the NW), the first hexagonal ML (here and next frames:

horizontal arrow) started propagating by step flow (inclined arrow). At 6.25 s, after nucleation of a second ML, the first two MLs clearly appear in hexagonal stacking (vertical yellow line) over the ZB segment (tilted line). **d** Frame extracted from Supporting Video SV2 showing an example of crystal phase heterostructure suitable for the fabrication of crystal phase qubits[10]. Scale bar: 5 nm.

The effect of the quasi-instantaneous change of contact angle is demonstrated in Fig. 2, which shows a NW growing while the bias is alternated between 0 and 60 V. Through repeated application of the E-field (Fig. 2a, b), multiple WZ insertions are synthetized. When the E-field is turned on and off, the contact angle is instantaneously modified (within the camera time resolution of 0.25 s) from well below to well above the critical angle of 125° identified in previous studies. Simultaneously, the crystal phase undergoes an abrupt modification.

This is further demonstrated by a frame-by-frame analysis of a single transition (Fig. 2c). At time 5.5 s, in absence of E-field, the NW is growing in the ZB phase. The E-field is switched on between 5.5 s and 5.75 s. This second frame thus captures the faint image of the NW before applying the E-field (which slightly shifts the NW and electron beam for imaging) and, in the stronger image at left, the yet incomplete first ML that has nucleated after switching. Close examination shows that this ML (which is nearly complete at 6 s) is an h-type ML, i.e. in hexagonal stacking position A if the last two ZB MLs are labelled AB. This is more clearly seen at 6.25 s, with a second ML propagating, and both MLs in h position (A and B, respectively). Forming a first h ML is a necessary condition for WZ growth and we demonstrate that this is correlated with applying the E-field. To grow a WZ sequence, we maintain the E-field for some time (Fig. 2a, b). However, if required (for instance to produce isolated planar faults or axial arrays thereof), we might also grow a single h ML by switching off the E-field immediately after its nucleation. Similar observations are made at the reverse transition. This confirms the robust correlation between contact angle and crystal phase achieved using our method. Therefore, complex crystal phase heterostructures can be synthetized with the lengths of ZB and WZ segments simply controlled by the duration of the E-field application.

As an example, we explored the possibility of fabricating a novel multiqubit platform based on CPQDs in NWs designed and theoretically studied by Li et al.[10]. The structure features two ZB segments separated by a WZ barrier in a WZ NW, thus requiring most of the NW to remain in the WZ phase. While Fig. 2a demonstrates the feasibility of synthesizing alternating ZB and WZ dots using the E-field, this raises the question of fabricating ZB dots in a WZ NW. Fig. 2d shows that it is indeed feasible. A WZ NW is grown by constant application of an E-field. Two distinct ZB dots are embedded in this NW by temporarily stopping the E-field. This confirms the capacity of the E-field method for controlled and reproducible growth of multiple CPQDs with definite phase, size, and spacing in a single NW.

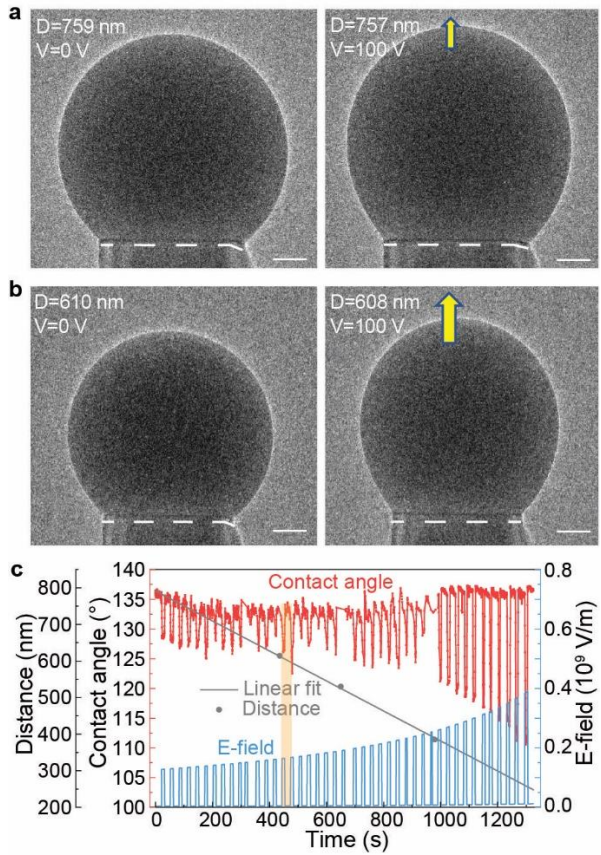

**Figure 3**: Behavior of the system upon decrease of the droplet- counter electrode distance D (see Fig. 1a). **a** Droplet at $D=D_1=759$ nm, with no bias (left) and with 100 V bias (right). **b** Droplet at shorter distance $D=D_2=610$ nm with no bias (left) and with bias of 100 V bias (right). In a, b, dashed lines are drawn slightly below the actual NW-droplet interface to guide the eye and the yellow arrows indicate the direction and strength of the E-field. Scale bars: 20 nm. **c** Measured distance between droplet tip and counter-electrode (dots) with linear fit as a function of time. The E-field is turned on and off repeatedly. Its nominal value (blue curve, right scale) is calculated by dividing the bias by the fitted distance. The contact angle (red curve) is an average between the values measured at the right and left sides of the droplet. At around 450 s (orange bar), the E-field becomes strong enough to suppress the corner truncation, which defines the nominal critical E-field. Missing data points result from temporary changes of the TEM magnification necessary to measure the droplet-electrode distance.

The droplet shape and contact angle are determined by the balance between two forces (the attraction of the conducting Ga-Au droplet carrying a non-uniform surface charge distribution by the biased counter electrode and the surface tension, which tends to minimize the droplet surface and thus keep it spherical). The surface tension of the droplet is an intrinsic physical quantity that is not expected to change here. Therefore, the amplitude of the E-field determines the extent of droplet deformation. To induce WZ formation, the E-field must therefore be strong enough to reduce the contact angle significantly. We first determine experimentally the minimum E-field that induces the phase change (termed below 'critical E-field') for a given NW diameter and then use finite element simulations to extend these results to a generic NW. To find the critical E-field, we simply keep applying the same bias while the growing NW approaches the counter-electrode. This results in an increasing E-field at the tip (Fig. 3). In practice, the electric bias is applied by short pulses of 100 V in order to avoid the change of NW diameter that would be induced by a constant bias[22]. Fig. 3c shows that when the E-field is off, the contact angle remains quasi-constant at about 135° (within ±5°) during the experiment. When the E-field is applied, the contact angle decreases to a value between 130°, when the NW is far from the counter-electrode, and 110° when the NW approaches it.

By analyzing the video of the NW growth, we find significant truncation at the TPL at any distance when the E-field is absent and, under bias, when the droplet is far away from the counter-electrode (Fig. 3a). In both cases, the contact angle exceeds the standard critical value of 125°. As the NW grows closer to the counter-electrode, a reduction in the truncation size becomes evident. At around 450 s, the critical E-field value is reached, after which the truncation is suppressed, and the NW-liquid interface becomes flat (Fig. 3b and Fig. S2 for details). For this NW, the critical E-field corresponds to a contact angle of about 125° and a distance about 600 nm between the tip of the droplet and the counter-electrode. The data for the first 1000 s is noisy partly because the NW sidewall morphology changes as the NW grows, which slightly alters droplet's contact line with NW stem. However, the trend of decreasing truncation size and contact angle is clear. The use of low magnification minimizes the total electron dose delivered to the nanowire and thus reduces the risk of beam-induced artifacts during extended imaging, and also enables periodic, rapid checks of the nanowire–electrode distance, allowing for accurate evaluation of the E-field.

At this point, the nominal E-field (calculated by dividing the bias by the NW-electrode distance) is around $1.6 \times 10^8$ V/m. We emphasize that this value pertains only to this NW and is not expected to be universal. This experiment is conducted at low magnification to minimize possible beam-induced artifacts as continuous observation of the NW is required, unlike for experiments in Fig. 1 and 2. We cannot identify the crystal phase, but as the E-field strength increases, the TPL truncation vanishes gradually, which in all our experiments conducted with atomic resolution, is a signature of WZ formation.

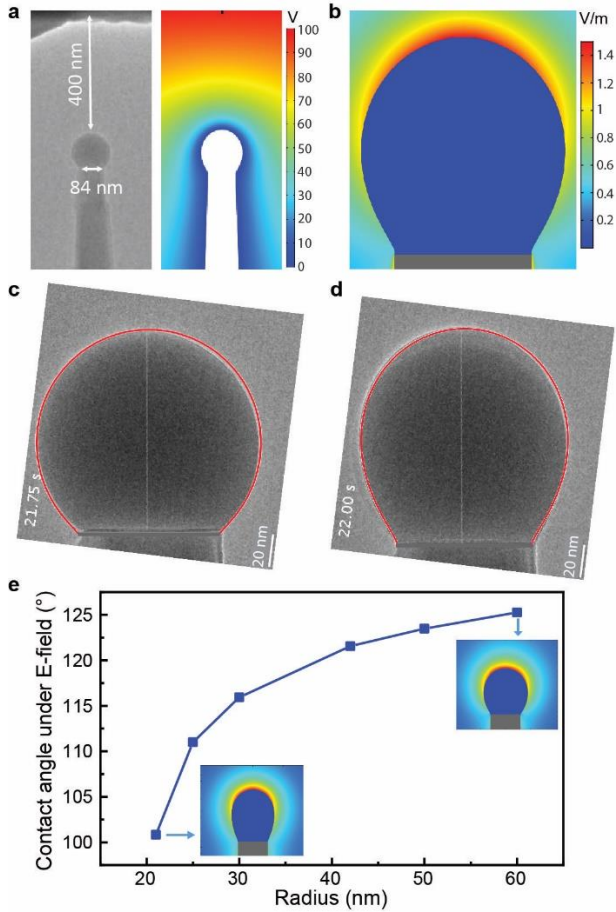

**Figure 4:** Simulation of the effect of the E-field on the droplet shape. **a** Left**:** Same NW and droplet as in Fig. 3. The counter-electrode distance, NW radii and contact angle are used as input for the simulation. Right: Simulation of the potential distribution, with bias of 100 V. The liquid-vapor surface tension used here is 0.75 J/m$^2$. **b** Droplet stretched in the presence of E-field (bias 100 V) and distribution of E-field amplitude. **c, d** Simulated droplet outline (red line) superimposed on the experimental image without (c) and with (d) E-field. The vertical white line shows the simulated droplet height. **e** Simulated contact angle when an E-field is applied for NWs of different radii, all parameters are the same except radius.

To extend our results to a wide range of NW radii and to provide guidance for future experiments, we simulate the effect of the E-field on the shape of the droplet using 2D and 3D finite element computation (details in Supporting Information). We first test our model by using distance, NW diameter and contact angle from the experiment of Fig. 3 in the simulation (Fig. 4a-4d). Fig. 4b demonstrates that the droplet tip experiences the strongest E-field, which is actually much larger than the nominal E-field, mainly due to the curvature of the droplet. Fig. 4c, d show that the simulated droplet shapes closely match the experiment. The differences between simulated and experimental height and contact angle are less than 1%. Such simulations offer easy access to the deformation of a much wider range of NW and droplet sizes than can be explored experimentally.

We can, for instance, investigate the impact of the NW radius on droplet deformation under E-field, at fixed counter-electrode distance and 0-bias contact angle. (Fig. 4e). For radii ranging from 25 nm to 60 nm, we find that smaller droplets undergo a larger deformation; namely, their contact angle under the E-field is smaller compared with bigger droplets. However, applications of CPQDs may impose a certain range of diameters[10,28]. Our simulations may then be used to find a suitable compromise.

In conclusion, we demonstrate the feasibility of applying an E-field to switch the crystal phase of GaAs NWs during growth. Our experiments suggest that the phase change mechanism remains similar to that observed in the absence of an applied E-field, re-emphasizing the crucial role of the contact angle. The application of the electric field alters both the droplet height and contact angle, and these changes closely correlate with crystal phase switching. By repeatedly switching the E-field on and off, heterostructures containing WZ or ZB QDs embedded in a ZB or WZ NW are successfully fabricated. We find that the E-field must exceed a radius-dependent critical value to deform the droplet sufficiently and induce phase changes. The experimental deformations are well reproduced by simulations, which also predict that narrower NWs deform more than wider ones under a given E-field. Our E-field method could be applied to other VLS-grown NWs and would provide great flexibility in the fabrication of heterostructures. It thus paves the way for the development of crystal phase quantum devices.

**Methods**

The cantilevers were fabricated in a process similar to that described by Kallesoe et al.[23]. The silicon device layer is 4±0.5 μm thick, with the cantilever sidewalls being the desired {111} planes. Two cantilevers are separated by a 2 μm gap in which NWs are grown. For growth, the cantilevers are resistively heated to around 550°C by running a typical current of 7 mA under 8 V through one cantilever loop. This is supplied by a Keithley 2401 Source Measure Unit operated as a constant voltage source. The temperature is pre-calibrated using Raman spectroscopy in a vacuum chamber. A second power supply controls the electric bias between the two cantilevers, which deforms the droplet. Up to the maximum voltage applied (100 V), the leakage current remains negligible (below 1 μA).

We observe the growth of GaAs NWs *in situ* using NanoMAX, a specially modified Cs corrected Titan environmental TEM[16] where total metalorganic precursors could be introduced up to a pressure of $10^{-2}$ mbar. Trimethylgallium (TMGa, typical partial pressure $3\times10^{-5}$ mbar) and tert-butylarsine (TBAs, typical partial pressure $6\times10^{-4}$ mbar) are used as precursor gases, without carrier gas. The Au catalyst of nominal thickness around 1 nm is deposited by sputtering.

Experiments are typically carried out at a growth rate between 0.5 and 4 ML/s and high-resolution movies are recorded using a Gatan US1000 camera at a rate of 4 frames per second. The images are analyzed using manual measurement by ImageJ software and an automated script that determines the relevant parameters, such as NW diameter, volume and contact angle of the liquid catalyst droplet.

For COMSOL simulation details, see ref. 26.

## ASSOCIATED CONTENT

**Supporting Information**

Supporting Information is available free of charge on the ACS Publications website upon reasonable request.

## AUTHOR INFORMATION

**Corresponding Author**

*E-mail: federico.panciera@cnrs.fr (F.P.), mirsaidov@nus.edu.sg (U. M.)

**Acknowledgements**

This research was funded by the French National Research Agency, ANR, through the TEMPOS Equipex NanoMAX, grant number ANR-10-EQPX-50, and the ELEPHANT project, grant number ANR-21-CE30-0012-01. French network of large high-end facilities (RENATECH) is also acknowledged for the fabrication process. ETEM experiments were carried out with the NanoMAX instrument at the Centre Interdisciplinaire de Microscopie Electronique de l'Ecole polytechnique (CIMEX), which is gratefully acknowledged. The authors thank Frances Ross and Kristian Mølhave for insightful discussions that contributed to this work.